\documentclass[aps,showpacs,prb,two column,floatfix,showpacs]{revtex4}
\usepackage{amsmath,amssymb,graphicx,bm}

\usepackage[english]{babel}
\usepackage[utf8]{inputenc}
\usepackage[a4paper]{geometry}
\newcommand{\Imm}{\mathop{\rm Im}\nolimits}

\newcommand{\wofz}{\mathop{\rm w}\nolimits}

\newcommand{\erfc}{\mathop{\rm erfc}\nolimits}

\newcommand{\sgn}{\mathop{\rm sgn}\nolimits}
\begin{document}

\title{Critical metal-insulator transition and divergence in a two-particle irreducible vertex in disordered and interacting electron systems}

\author{V.  Jani\v{s}}\email{janis@fzu.cz} \author{V. Pokorn\'y}\email{pokornyv@fzu.cz} 
\affiliation{Institute of Physics, Academy of Sciences of the Czech
  Republic, Na Slovance 2, CZ-18221 Praha 8, Czech Republic}

\date{\today}


\begin{abstract}
We use the dynamical mean-field approximation to study singularities in the self-energy and a two-particle irreducible vertex induced by the metal-insulator transition of the disordered Falicov-Kimball model. We set general conditions for the existence of a critical metal-insulator transition caused by a divergence of the imaginary part of the self-energy. We calculate explicitly the critical behavior of the self-energy for the symmetric and asymmetric disorder distributions. We demonstrate that the metal-insulator transition is preceded by a pole in a two-particle irreducible vertex. We show that unlike the singularity in the self-energy the divergence in the irreducible vertex does not lead to non-analyticities in measurable physical quantities. We reveal universal features of the critical metal-insulator transition that are transferable also to the Mott-Hubbard transition in the models of the local Fermi liquid.      
\end{abstract}

\maketitle 

\section{Introduction}
%
%
%

The electron correlations and the configurational disorder,  although different in the origin and in the way they manifest themselves, have in common that they hinder free propagation of electrons in metals. When sufficiently strong they both may lead to a metal-insulator transition. There are two types of the metal-insulator transition at zero temperature. The first one is the so-called Mott-Hubbard transition at which the density of electrons on the Fermi energy vanishes. The second one is the Anderson localization transition at which the electron diffusion vanishes but the density of states at the Fermi energy remains nonzero. The latter transition is, however,  driven entirely by the configurational disorder and will not be addressed in this paper.          

The early ideas  that electron correlations in chemically homogeneous materials may lead to a metal-insulator transition came from Mott.\cite{Mott49} First quantitative scenarios of a correlation-induced metal-insulator transition were proposed by Hubbard and Gutzwiller. The former picture is based on a self-consistent theory of conduction electrons effectively scattered on other thermally equilibrated localized electrons with the opposite spin. Multiple scatterings on the localized electrons then lead to repulsion of the electron states from the Fermi energy and eventually to a metal-insulator transition.\cite{Hubbard64} The latter approach uses a variational method in which electron correlations reduce the width of the otherwise undeformed energy band. A metal-insulator transition is then signaled by vanishing of the bandwidth.\cite{Gutzwiller63}  The two scenarios for the correlation-driven metal-insulator transition are non-critical, all characteristic parameters are continuous and bounded at the transition. There is no divergence in the spectral or response functions. They hence describe a transition from a metal to a band insulator but not to a Mott insulator where the self-energy has a pole at the Fermi energy.

Neither of these two pictures reflects reality, since they both completely neglect dynamical fluctuations that interfere these static scenarios by the Kondo effect present in the local Fermi liquid.\cite{Hewson93} A reliable theory of the correlation-induced metal-insulator transition without any symmetry breaking in Fermi liquids must hence cope with this dynamical effect. A resolution to the problem of the correlation-induced metal-insulator transition was offered for the Hubbard model a few decades ago by the dynamical mean-field theory (DMFT).\cite{Georges96,Imada98} This approximation becomes exact in infinite spatial dimensions and contains the Kondo asymptotics of the single-impurity Anderson model. But in addition to the Kondo physics it contains a self-consistency between the impurity (local Fermi liquid) and the surrounding metallic environment that can suppress the Kondo peak and enforce an insulating state. There is, however, only numerical evidence for the existence of a metal-insulator transition within the  dynamical mean-field theory and a comprehensive analytic solution is still missing.\cite{Zhang93,Bulla99}

Unlike the Hubbard model, that cannot be solved analytically, a simpler Falicov-Kimball model (FKM) allows for an analytic solution in infinite dimensions.  Its infinite-dimensional solution resembles that of the Hubbard approximation for the metal-insulator transition, called Hubbard III.\cite{Janis90,Janis93a} The Falicov-Kimball model is not a Fermi liquid and it is free of the Kondo effect. It was explicitly shown that it displays a metal-insulator transition on a Bethe lattice.\cite{vanDongen97} The transition at half filling is critical with a divergence of the self-energy.\cite{Freericks03} Even in this model, however,  a detailed critical asymptotics of the self-energy and the spectral functions at this metal-insulator transition is missing.

A metal-insulator transition, similar in properties to the Mott-Hubbard transition, is known also in disordered systems where it is called a split-band transition. It is described reliably by the coherent-potential, mean-field, approximation.\cite{Elliot74} A similarity between the Mott-Hubbard and the metal-insulator transition in a binary alloy was unveiled in Ref.~\onlinecite{Velicky68}.  The split-band transition is generally not critical, it means there are no divergences at the transition point. It occurs when the chemical potential reaches one of the edges of the renormalized band at which the imaginary part of the self-energy vanishes. The  exception to it is the split-band transition for the symmetric distributions of the random potential. In this case the transition is marked by a divergence of the self-energy. This is the feature that is shared with the Mott-Hubbard transition. It is hence apparent that a unified and universal description of the Mott-Hubbard and the metal-insulator transition in random systems with the diverging self-energy should exist.                  

The Mott-Hubbard transition can be fully described with one-electron functions, the self-energy and the local Green function. Generally, continuous phase transitions are marked by singularities in response functions, that is, two-particle quantities.  It is argued that the local spin susceptibility diverges at the Mott-Hubbard transition.\cite{Georges96} There is no two-particle function that would diverge at the metal-insulator transition in the disordered Falicov-Kimball model. There is, however, a singularity in a two-particle irreducible vertex prior the metal-insulator transition is reached.\cite{Janis01a,Janis01b}  This singularity is not, however, connected with a symmetry breaking, a long-range order, or a non-analytic behavior of measurable quantities. It is an indicator of proximity of a metal-insulator transition as discussed recently in the Hubbard as well as the Falicov-Kimball models.\cite{Schafer13} The exact quantitative relation between the singularity in the two-particle irreducible vertex and the singularity of the self-energy at the metal-insulator transition has not yet been demonstrated.   

The aim of this paper is to find the explicit low-frequency asymptotics of spectral functions at the metal-insulator transition characterized by the diverging imaginary part of the self-energy and to clarify its relation to a singularity in the two-particle irreducible vertex in the disordered Falicov-Kimball  model. This model combines the effects of electron correlations and disorder and allows for a solution in the mean-field limit, exact solution on an infinite-dimensional lattice. We set the conditions for the existence of this type of the metal-insulator transition for general input parameters, such as the density of states of the conduction electrons, band filling, and the type of the disorder distribution. We further demonstrate that the singularity in the two-particle irreducible vertex always precedes the divergence in the self-energy. Although we study a model that is not a Fermi liquid, our analysis unveils universal features of the metal-insulator transition with the diverging self-energy that are transferable also on the Mott-Hubbard transition, if it exists, in models of the strongly correlated local Fermi liquid.

\section{Model and a mean-field solution}

The Falicov-Kimball model describes noninteracting electrons being scattered on a static, but spatially inhomogeneous atomic potential of ions forming a regular crystalline lattice. We consider the basic version of the spinless single-orbital Falicov-Kimball model  and take into account only one type
of the extended (conduction)  and one type of the localized electrons. We further assume
homogeneity in the distribution of the localized electrons and hence
the noninteracting part of the Hamiltonian describing such a situation
reads
\begin{subequations}\label{eq:FK-Hamiltonian}
  \begin{align}
    \widehat{H}_0 &= \sum_{\mathbf{k}}\epsilon(\mathbf{k})
    c^{\dagger}(\mathbf {k}) c(\mathbf{ k}) + E_f \sum_i f^{\dagger}_i
    f^{\phantom{\dagger}}_i \ .
  \end{align}%
  We are interested in the dynamical properties of the delocalized
  electrons induced by the fluctuations of the atomic potential the
  conduction electrons feel. To this purpose we introduce an interacting
  term
  \begin{align}\label{eq:FK-Hint}
   \widehat{H}_I & = \sum_i \left[V_i + U f^{\dagger}_i
      f^{\phantom{\dagger}}_i \right] c^{\dagger}_{i}
    c^{\phantom{\dagger}}_{i}
  \end{align}\end{subequations}
where $c_i = N^{-1}\sum_{\mathbf{k}}c(\mathbf{k}) \exp
\{-i\mathbf{k}\cdot \mathbf{R}_i\}$. We denoted $V_i$ the atomic level of
the ion situated in the elementary cell centered around the lattice vector
$\mathbf{R}_i$, and $U$ is the interaction strength between the
conduction and the localized electrons. We generally assume that the atomic
potential $V_i$ is a random variable with a static, site-independent
probability distribution with its values determined externally.  If
$U=0$ the full Hamiltonian $\widehat{H} = \widehat{H}_0 +
\widehat{H}_I$ is that of the Anderson model with disordered electrons
and if $V_i=0$, the full Hamiltonian describes the homogeneous Falicov-Kimball
model.  Actually, both the contributions to the Hamiltonian from
Eq.~\eqref{eq:FK-Hint} introduce a randomness into the
distribution of the atomic energy levels the conduction electrons
feel. Potential $V_i$ represents a static (quenched) randomness and
interaction $U$ a dynamical (annealed) one. Both contributions can
be treated on the same footing.

An exact solution to FKM is known in the
limit of infinite spatial dimensions. The equilibrium thermodynamics of the disordered FKM in $d =\infty$  was analyzed in Ref.~\onlinecite{Janis92}. The functional
of the averaged grand potential was found to be represented via a set
of complex variational parameters $G_n$ and $\Sigma_n$, where
index $n$ corresponds to the $n$th fermionic Matsubara frequency,
\begin{multline}\label{eq:FE-infty} \beta 
    \Omega  = -
  \left\langle \ln \left[1 + \exp\left\{\beta(\mu - E_f
        -\mathcal{E}_V)\right\} \right]\right\rangle_{av} \\ - \sum_{n=-\infty}^\infty\left\{
    \int_{-\infty}^\infty d E \rho_{0}(E) \ln \left[i\omega_n + \mu - E -
      \Sigma_n\right]\right. \\ \left. + \left\langle \ln \left[1 +
        G_n(\Sigma_n - V) \right]\right\rangle_{av}\right.\bigg\} \
  . \end{multline}%
Symbol $\left\langle X(V) \right\rangle_{av} = \int dV P(V) X(V)$ stands for averaging
over the distribution of the random potential $V_i$ with a site-independent probability distribution $P(V)$.  The density of states of the conduction electrons of the pure and non-interacting model was denoted $\rho_{0}$. The shift of the $f$-electron atomic level $\mathcal{E}_V$ is determined via the same complex numbers $G_n$ and $\Sigma_n$
\begin{equation}\label{eq:E-shift}
  \mathcal{E}_V = - \frac 1\beta\sum_{n=-\infty}^\infty  \ln \left[1 - \frac {U
      G_n}{1 + G_n(\Sigma_n -V)}\right]
\end{equation}
and depends on the actual value of the random atomic potential $V$.
 
The equilibrium thermodynamics is obtained as a stationarity point  of the averaged grand potential $\Omega$ from Eq.~\eqref{eq:FE-infty}
with respect to small variations of complex numbers $\Sigma_n$ and
$G_n$.  Vanishing
of variations of the former and the latter parameters leads to a pair
of equations for each Matsubara frequency $\omega_n$
%
  \begin{align}\label{eq:1PGF}
   & G_n = \int_{-\infty}^\infty \frac{ d
      \epsilon\rho_{0}(\epsilon)}{i\omega_n + \mu - \epsilon -\Sigma_n}\ ,
\end{align} 
\begin{multline}   
    \label{eq:SE} 
    1 = \left\langle \frac{1 - f(E_f +
        \mathcal{E}_V)}{1 + G_n(\Sigma_n - V)} \right. \\ \left. + \frac{f(E_f +
        \mathcal{E}_V)}{1 + G_n(\Sigma_n - V - U)}\right\rangle_{av} \ ,
  \end{multline} 
where we denoted the Fermi function $f(x)= 1/[1 + \exp\{\beta( x - \mu)\}]$.
The first equation states that  $G_n$ in equilibrium is the local
element of the one-electron thermal Green function with a self-energy
$\Sigma_n$. The second equation determines the value of the
equilibrium self-energy. These equations of the thermal equilibrium must
be completed with an equation determining the chemical potential $\mu$
from the total electron density $n$. This equation then is
\begin{subequations}\label{eq:n-def}
\begin{align}\label{eq:ntot-def}
n &= n_{f} + n_{c}\ 
\end{align}
with partial particle densities
\begin{align}
\label{eq:ncf-def}
  n_{f}& = \left\langle f(E_f + \mathcal{E}_V)\right\rangle_{av}\ , \\ 
   n_{c} & = \frac 1 \beta
  \sum_{n=-\infty}^\infty G_n e^{i\omega_n0^+} \ . \label{eq:nc-def}
\end{align}
\end{subequations}
Equations \eqref{eq:E-shift}-\eqref{eq:n-def} fully determine the
equilibrium thermodynamics for a given temperature $T = 1/\beta$ and a total
particle density $n$.

Only one-particle equilibrium functions can be directly calculated
from the grand potential $\Omega$. The thermodynamic potential contains the full information to describe the behavior near the equilibrium metal-insulator transition. To derive phase transitions with a symmetry breaking one needs to evaluate higher-order correlation functions for which we have to slightly perturb equilibrium and look at the corresponding non-local response functions.\cite{Janis05b} This essentially goes beyond the standard mean field theory (limit to infinite lattice dimensions) where only integrals with the density of states are enough to know.\cite{Janis99} It is important that we do not need higher-order and non-local Green functions in the study of the metal-insulator transition.

\section{Critical metal-insulator transition}

We analyze the behavior of the retarded local one-electron Green function of the conduction electrons from the defining equation
\begin{align}
\label{DFKM:DysonEquation}
G^R(\omega) &=\int_{-\infty}^{\infty} \frac{d\epsilon\rho_0(\epsilon)}
{\omega+i0^+  +\mu - \epsilon - \Sigma^R(\omega)}\ .
\end{align}
Self-energy $\Sigma^{R}(\omega)$ in this representation, containing the effects of the Coulomb repulsion and of the random atomic potential, is determined in the disordered Falicov-Kimball model from the Soven equation
\begin{align}\label{eq:DFKM-SovenEquation}
1&=\left\langle\frac{1-n_f}{1+G^R(\omega)[\Sigma^R(\omega)-V]} \right. \nonumber \\ 
& \left. \qquad+\frac{n_f}{1+G^R(\omega)[\Sigma^R(\omega)-V-U]}\right\rangle_{av},
\end{align}
The configurationally-dependent density of the local electrons is calculated from Eq.~\eqref{eq:ncf-def} where 
\begin{align}
\mathcal{E}_{V } & =  - \int_{-\infty}^{\infty}\!\! \frac{d\omega}\pi f(\omega) \nonumber \\
&\quad  \times \Im  \ln \left[1 - \frac {U G^{R}(\omega)}{1 + G^{R}(\omega)\left(\Sigma^{R}(\omega) - V\right)}\right] \ .
\end{align}
The density of the conduction electrons, or band filling then is  
\begin{align}
n_{c} & = - \int_{-\infty}^{\infty}\!\! \frac{d\omega}\pi f(\omega) \Im
  G^R\left(\omega\right) \ .
  \end{align}

We solve  Eq.~\eqref{eq:DFKM-SovenEquation} asymptotically in the critical region of the metal-insulator transition with the diverging self-energy. If we denote $\Sigma^{R}(\omega) = R(\omega)  - i S(\omega)$,  then $S(\omega) \ge 0$. The critical point is signaled by a divergence $S(\omega) \to \infty$ for $\omega \to \omega_{0}$, where $\omega_{0}$ is the frequency at which the self-energy diverges. If $\omega_{0} = 0$ then the critical point occurs at the Fermi energy. For the asymmetric disorder distribution or away from half filling the self-energy may diverge away from the Fermi energy, $\omega_{0} \neq 0$. 

It is convenient to introduce small parameters in the critical region of the metal-insulator transition. We denote  $r = R/(R^{2} + S^{2})$, $s = S/(R^{2} + S^{2})$ and suppress frequency dependence in these functions.  We rewrite the retarded propagator in these variables 
\begin{multline}
G^{R}(\omega) = \int_{-\infty}^{\infty}\frac{d \epsilon\rho_{0}(\epsilon)}{\omega + \mu - \epsilon - R + iS} \\ = - (r + i s)\int_{-\infty}^{\infty}\frac{d \epsilon\rho_{0}(\epsilon)}{ 1 - (r + i s)(\omega + \mu - \epsilon)} \ .
\end{multline}
To make the expected small parameters dimensionless  we introduce a new energy unit $\sqrt{\left\langle\overline{\epsilon}^{2}\right\rangle}$ and turn all the energy-dependent variables dimensionless. We further denote $\langle\epsilon^{n}\rangle  = \int d\epsilon \rho_{0}(\epsilon) \epsilon^{n}$ and $\overline{\epsilon} = \epsilon  - \langle \epsilon\rangle$.  We analogously introduce a new frequency variable  $\overline{\omega} = \omega + \mu - \langle \epsilon\rangle$.

Instead of the Green function we will use the following two functions in the Soven equation
\begin{align}\label{eq:g0}
\gamma_{0} &= \int_{-\infty}^{\infty}\frac{d \epsilon\rho_{0}(\epsilon)}{ 1 - (r + i s)(\overline{\omega} - \overline{\epsilon})} \ , \\ \label{eq:ge}
\gamma_{\epsilon} &= \int_{-\infty}^{\infty}\frac{d \epsilon \rho_{0}(\epsilon)\overline{\epsilon}}{ 1 - (r + i s)(\overline{\omega} - \overline{\epsilon})} \ .
\end{align}
With these functions we obtain 
\begin{multline}
 1 + G^{R}(\omega)(\Sigma^{R}(\omega) - V) \\ = - (r + i s)  \left[(\overline{\omega} - V)\gamma_{0} - \gamma_{\epsilon}\right] \\ = - (r + i s) \left[(\overline{\omega} - V) X + i s Y\right]
\end{multline}
where  we introduced real functions  $X,Y$. They will be expanded in small parameters $r$ and $s$. To do so we rewrite the Soven equation in this representation 
\begin{equation}\label{eq:Soven-complex}
r + i s =  - \left\langle \frac{1}{(\overline{\omega} - V) X + i s Y}\right\rangle_{V,T}\ ,
\end{equation}
with the real and the imaginary components
\begin{align}\label{eq:r-general}
r &= - \left\langle \frac{(\overline{\omega} - V) X}{(\overline{\omega} - V)^{2} X^{2} + s^{2} Y^{2}}\right\rangle_{V,T} \ ,\\
s & = s \left\langle \frac{Y}{(\overline{\omega} - V)^{2} X^{2} + s^{2} Y^{2}}\right\rangle_{V,T}\ . \label{eq:s-general}
\end{align}
We denoted the thermal and configurational averaging by a single symbol $\langle \cdot\rangle_{V,T}$. It acts on any function of the random potential $V$ as follows 
$$
\left\langle F(V)\right\rangle_{V,T} = (1 - n_{f}) \left\langle F(V) \right\rangle_{av} + n_{f}\left\langle F(V + U) \right\rangle_{av} \ . 
$$ 
The temperature dependence enters only via the density of the local $f$-electrons $n_{f}$, Eq.~\eqref{eq:ncf-def}. Such an extended averaging, merging thermal and configurational fluctuations, is possible to introduce, since the density of the local $f$-electrons $n_{f}$ is a conserving quantity. 

We expand the two real functions $X,Y$ in small parameters $r,s$. We will need  to expand them up to third order in $r$ and $s$ 
\begin{multline}
X \doteq 1 - r \left[ \frac{1}{ V - \overline{\omega}} - \overline{\omega}\right] \\  -  \left( r^{2} - s^{2}\right)\left[\frac{2\overline{\omega}}{V - \overline{\omega}} - \overline{\omega}^{2} - 1 \right] \\
- r\left( r^{2}-3s^{2}\right)\left[\frac{\left\langle\overline{\epsilon}^{4}\right\rangle + 3 \overline{\omega}^{2} }{ V - \overline{\omega}} - \overline{\omega}\left( \overline{\omega}^{2} + 3 \right)  \right]
\end{multline}
and
\begin{multline}
\frac{Y}{V - \overline{\omega}} \doteq \frac{1}{ V - \overline{\omega}} - \overline{\omega}\\  + 2r \left[\frac{2\overline{\omega}}{ V - \overline{\omega}} - \overline{\omega}^{2} - 1 \right] 
\\
+ \left(3r^{2} - s^{2}\right) \left[\frac{\left\langle \overline{\epsilon}^{4}\right\rangle + 3 \overline{\omega}^{2} }{V - \overline{\omega}} - \overline{\omega}(\overline{\omega}^{2} + 3)  \right]\ .
\end{multline}

The leading asymptotic order results in equations 
\begin{align}\label{eq:MIT-Rzero}
0 &=  \left\langle \frac{1}{V - \overline{\omega}_{0}}\right\rangle_{V,T}\ , \\
s & = s \  \left\langle \frac{1}{(V - \overline{\omega}_{0})^{2}}\right\rangle_{V,T}  \ . \label{eq:MIT-Izero}
\end{align}
The first equation defines the frequency (energy) $ \omega_{0} = \overline{\omega}_{0} - \mu + \langle \epsilon\rangle$ at which the self-energy diverges and the second one determines the critical value of the disorder (interaction) strength.  Note that we did not need to use finiteness of the bandwidth of the conduction electrons to produce a diverging self-energy and vanishing of the spectral function. We show later on that only for finite bandwidths of the conduction electrons the diverging self-energy leads to opening of a gap in the energy spectrum. In this respect the Mott-Hubbard and the critical metal-insulator transition represent the same universal phenomenon independent of the detailed structure of the dispersion relation and of the density of states of the conduction electrons. 

In the next step we determine the leading frequency dependence of small parameters $r(\overline{\omega})$ and $s(\overline{\omega})^{2}$. We obtain in the leading asymptotic order $\overline{\omega} \to \overline{\omega}_{0}$ a set of coupled equations 
\begin{widetext}
\begin{multline}\label{eq:MIT-Re}
r(\overline{\omega})^{3}\left[V_{4}(\overline{\omega}) + \left( \left\langle\overline{\epsilon}^{4}\right\rangle  - 2\right) V_{2}(\overline{\omega}) \right] + r(\overline{\omega})^{2}V_{3}(\overline{\omega})  \\ 
+ r(\overline{\omega})\left[V_{2}(\overline{\omega}) - 1 - 3 s(\overline{\omega})^{2}\left(V_{4}(\overline{\omega}) + \left(\left\langle\overline{\epsilon}^{4}\right\rangle  - 2 \right)V_{2}(\overline{\omega}) \right)\right]
+ V_{1}(\overline{\omega}) =0
\end{multline}
and 
\begin{multline}\label{eq:MIT-Im}
s(\overline{\omega}) \left\{ V_{2}(\overline{\omega})  - 1  + 2 r(\overline{\omega})  V_{3}(\overline{\omega}) + 3 r(\overline{\omega})^{2} \left[V_{4}(\overline{\omega})  + \left( \left\langle\overline{\epsilon}^{4}\right\rangle  - 2\right) V_{2}(\overline{\omega}) \right] \right. \\  \left.  - s(\overline{\omega})^{2}\left[ V_{4}(\overline{\omega}) + \left(\left\langle\overline{\epsilon}^{4}\right\rangle  - 2 \right)V_{2}(\overline{\omega})\right]\right\} = 0  
\end{multline}
where we abbreviated  $V_{n}(z) = \left\langle (V - z)^{-n}\right\rangle_{V,T}$.


The two equations can be decoupled in that we resolve function $s(\overline{\omega})^{2}$. They have two solutions. If $V_{2}(\overline{\omega}_{0}) <  1$, the strong-scattering limit (insulating state), then $s(\overline{\omega}) = 0$ and  
\begin{equation}\label{eq:r-ins}
r(\overline{\omega})^{3}\left[V_{4}(\overline{\omega}) + \left( \left\langle\overline{\epsilon}^{4}\right\rangle  - 2\right) V_{2}(\overline{\omega}) \right] + r(\overline{\omega})^{2}V_{3}(\overline{\omega}) - r(\overline{\omega})\left[ 1 -V_{2}(\overline{\omega})\right] + V_{1}(\overline{\omega}) =0 \ .
\end{equation}
In the weak-scattering limit (metallic state), $V_{2}(\overline{\omega}_{0}) \ge  1$,  we obtain 
\begin{equation}\label{eq:r-met}
8 r(\overline{\omega})^{3}\left[V_{4}(\overline{\omega}) + \left( \left\langle\overline{\epsilon}^{4}\right\rangle  - 2\right) V_{2}(\overline{\omega}) \right] + 5 r(\overline{\omega})^{2}V_{3}(\overline{\omega}) + 2 r(\overline{\omega})\left[ V_{2}(\overline{\omega}) - 1\right] - V_{1}(\overline{\omega}) =0 \ 
\end{equation}
\end{widetext}
and 
\begin{align}\label{eq:s-met}
s(\overline{\omega}) ^{2} & = \frac{V_{2}(\overline{\omega}) - 1 + 2r(\overline{\omega})V_{3}(\overline{\omega}) }{V_{4}(\overline{\omega}) + \left( \left\langle\overline{\epsilon}^{4}\right\rangle  - 2\right) V_{2}(\overline{\omega})} + 3 r(\overline{\omega})^{2}\ .
\end{align}

The solution in the strong-scattering limit extends to a finite gap only if the bandwidth of the conduction electrons is finite. When the conduction band has infinite tails we have to correct the expansion in $z= r + i s$ and to add a contribution from the pole in the one-electron propagator. We then have the following representation for the denominator of the right-hand side of Eq.~\eqref{eq:Soven-complex}
\begin{multline}\label{eq:exp-tails}
(\overline{\omega} - V) \gamma_{0} - \gamma_{\epsilon}= \frac{i\pi \rho_{0}(\omega + \mu - 1/r)}{z^{2}}\left( 1 - z V\right) \\  + \sum_{n=0}^{\infty} z^{n}\left\langle (\overline{\omega} - \epsilon - V)(\overline{\omega} - \epsilon)^{n}\right\rangle \ .
\end{multline}
We can see from this formula that the gap, $s(\overline{\omega})  = 0$,  opens in the strong-scattering regime only for $\rho_{0}(\omega + \mu - 1/r) = 0$, that is, if the band of the energy states of the conduction electrons is bounded and $\omega + \mu - 1/r$ lies outside its edges. Otherwise the self-energy remains nonzero due to long band tails of the density of states of the conduction electrons. In the insulating phase, where  $s(\overline{\omega}) = 0$ identically, the self-energy diverges for a frequency $\overline{\omega}_{0}$ obeying Eq.~\eqref{eq:MIT-Rzero}. The line of the singular real-part of the self energy in the energy gap starts at a band edge at which also the condition from Eq.~\eqref{eq:MIT-Izero} holds.  

The metallic phase offers a richer critical behavior. If $V_{2}(\overline{\omega}_{0}) >  1$ we obtain asymptotic solutions
\begin{align}
r(\overline{\omega}) & = \frac{V_{1}(\overline{\omega})}{2\left(V_{2}(\overline{\omega}_{0}) - 1 \right)} \ ,\\
s(\overline{\omega})^{2} & = \frac{\left(V_{2}(\overline{\omega}_{0}) - 1 \right)^{2} + V_{3}(\overline{\omega}) V_{1}(\overline{\omega})}{\left( V_{2}(\overline{\omega}) - 1\right) \left[V_{4}(\overline{\omega}) + \left( \left\langle\overline{\epsilon}^{4}\right\rangle  - 2\right) V_{2}(\overline{\omega}) \right]}\nonumber \\
& \quad + \frac{3V_{1}(\overline{\omega})^{2}}{4 \left( V_{2}(\overline{\omega}) - 1\right)^{2}} \ .
\end{align}
The self-energy does not diverge, it only reaches its maximum at $\overline{\omega}_{0}$ with the value
\begin{align}
s(\overline{\omega}_{0})^{2} & = \frac{ V_{2}(\overline{\omega}_{0}) - 1}{V_{4}(\overline{\omega}_{0}) + \left( \left\langle\overline{\epsilon}^{4}\right\rangle  - 2\right) V_{2}(\overline{\omega}_{0}) } \ . 
\end{align}
The self-energy diverges at a critical value of the scattering potential determined from the equation $V_{2}(\overline{\omega}_{0}) = 1$. Notice that the critical point can be reached only if $V_{4}(\overline{\omega}_{0}) +  \left\langle\overline{\epsilon}^{4}\right\rangle  - 2 > 0$, so that the solution $s(\overline{\omega})^{2} \searrow 0$  exists.  

The frequency asymptotics of the self-energy at the critical point depends on whether the probability distribution of the scattering potential is  symmetric or asymmetric. For the symmetric distribution, $V_{2n + 1}(\overline{\omega}_{0}) = 0$,  the two functions of interest asymptotically behave as 
\begin{align}
r(\overline{\omega})^{3} & = \frac{\Delta\overline{\omega}}{8 \left[V_{4}(\overline{\omega}_{0}) + \left\langle\overline{\epsilon}^{4}\right\rangle  - 2\right] } \ ,\\
s(\overline{\omega})^{6} & = \frac{27\Delta\overline{\omega}^{2}}{8 \left[V_{4}(\overline{\omega}_{0}) +  \left\langle\overline{\epsilon}^{4}\right\rangle  - 2\right] ^{2}}\ , 
\end{align}
where  we denoted $\Delta\overline{\omega} = \overline{\omega} - \overline{\omega}_{0}$.
Notice that the leading asymptotics of function $s(\overline{\omega})$ is fully determined by the asymptotics of function $r(\overline{\omega})$. Hence the real and imaginary parts of the  self-energy diverge with the same power, $\Delta\overline{\omega}^{-1/3}$.

In the case of an asymmetric distribution of the scattering potential,
$V_{2n+1}(\overline{\omega}_0) \ne 0$ for $n\ge 1$, we find that at the
critical point the self-energy $\Sigma^R(\omega)$ has metallic
character for one sign of $\omega-\omega_0$ (or of
$\Delta\overline{\omega}$) and insulating character for the opposite sign.
For $\Delta\overline{\omega} V_3(\overline{\omega}_0) > 0$, we obtain the
metallic solution  
\begin{align}
r(\overline{\omega})^{2} & = \frac{\Delta\overline{\omega}}{5 V_{3}(\overline{\omega}_{0}) } \ ,\\ \label{eq:s-off}
s(\overline{\omega})^{2} & = \frac 2{\sqrt{5}} \frac{\sqrt{\Delta\overline{\omega}V_{3}(\overline{\omega}_{0})}} {V_{4}(\overline{\omega}_{0}) +  \left\langle\overline{\epsilon}^{4}\right\rangle  - 2  }  \ .
\end{align}
If $\Delta\overline{\omega}V_{3}(\overline{\omega}_{0})  < 0$ then  we are in the insulating state with $s(\overline{\omega}) = 0$ and
\begin{align}
r(\overline{\omega})^{2} & =  - \frac{\Delta\overline{\omega}}{V_{3}(\overline{\omega}_{0}) } \ ,
\end{align}
since Eq.~\eqref{eq:r-ins} applies. It means that the critical metal-insulator transition with a diverging self-energy does not coincide with the split-band transition at which the solution is metallic from both sides, that is for $\Delta \overline{\omega}> 0$ as well as for  $\Delta \overline{\omega}< 0$.  The critical metal-insulator transition for the asymmetric distributions of the scattering potential hence coincides with a band edge. The critical behavior is asymmetric at this critical band edge. The imaginary part of the self-energy diverges as  $\Delta\overline{\omega}^{-1/4}$ when approaching the critical point from the metallic side  while the real part remains bounded. On the insulating side the imaginary part of the self-energy is zero and its real part diverges as  $\Delta\overline{\omega}^{-1/2}$ at the band edge. Both the real and the imaginary parts of the self-energy experience an infinite jump when crossing the critical band edge. 

As an example we calculate the critical metal-insulator transition for the homogeneous Falicov-Kimball model with a symmetric density of states, that is $V=0$ and $\left\langle \epsilon\right\rangle =0$. From Eq.~\eqref{eq:MIT-Rzero} and the definition of frequency $\overline{\omega} = \omega + \mu$ we obtain 
\begin{align}\label{eq:FKM-MIT1}
\omega_{0} &= (1 - n_{f})U - \mu\ .
\end{align} 
We use this solution in Eq.~\eqref{eq:MIT-Izero} and find the value for the critical interaction (in the energy units with $\sqrt{\left\langle\overline{\epsilon}^{2}\right\rangle} = 1$)
\begin{align}\label{eq:FKM-MIT2}
U_{c}^{2} & = \frac{1}{n_{f}(1 - n_{f})} \ . 
\end{align} 
Equations~\eqref{eq:FKM-MIT1} and~\eqref{eq:FKM-MIT2} determine a critical point at which the imaginary part of the self-energy diverges when approaching it from the metallic state. It is easy to evaluate the explicit critical asymptotics. The symmetric case is characterized by $n_{f} = 1/2$ and $\overline{\omega}_{0}= U/2$. Then the self-energy diverges as 
\begin{align}
\Re\Sigma(\omega) & \doteq \frac 2{7} \ \frac{\left(\left\langle \epsilon^{4}\right\rangle - 1\right)^{1/3}}{\left(\Delta\omega\right)^{1/3}} \ ,\\
\Im\Sigma^{R}(\omega) & \doteq  - \frac {2\sqrt{6}}{7} \ \frac{\left|\left\langle \epsilon^{4}\right\rangle - 1\right|^{1/3}}{\left|\Delta\omega\right|^{1/3}} \ , \label{eq:ImS-sym}
\end{align}  
where $\Delta\omega = \omega - U/2$, since $\mu = 0$ in this case. 

The critical behavior of the imaginary part of the self-energy on the metallic side ($\Delta\overline{\omega} > 0$) in the asymmetric situation then is for $n_{f} < 1/2$ 
\begin{align}
\Im\Sigma^{R}(\omega) & \doteq  - \left(\frac {5}{4 \Delta\overline{\omega}}\right)^{1/4} \nonumber \\ 
&\quad \times \frac{\left|1 + n_{f}(1 - n_{f})\left(\left\langle \epsilon^{4}\right\rangle - 5\right)\right|^{1/2}}{(1 - 2n_{f})^{1/4}\left[n_{f}(1- n_{f})\right]^{3/8}} \  .
\end{align} 
The real part of the self-energy from the metallic side acquires a finite value at the critical point  
\begin{align}
\left|\Re\Sigma(\omega_{0})\right| & \doteq  \frac{1 + n_{f}(1 - n_{f})\left(\left\langle \epsilon^{4}\right\rangle - 5\right)}{(1 - 2n_{f})^{1/2}\left[n_{f}(1- n_{f})\right]^{1/2}} \  .
\end{align} 
While it diverges in the insulating phase as 
\begin{align}
\left|\Re\Sigma(\omega)\right| & \doteq  \frac 1{|\Delta\overline{\omega}|^{1/2}}\ \frac{\left[n_{f}(1- n_{f})\right]^{1/2}}{(1 - 2n_{f})} \  .
\end{align} 
This singularity in the real part of the self-energy extends into the insulating state along the line determined from Eq.~\eqref{eq:MIT-Rzero}, that is, for $\omega_{0} = (1 - n_{f})U - \mu$. The divergence of the self-energy at $\overline{\omega}_{0}$ in the insulating phase is, however, linear for $\omega \to (1 - n_{f})U - \mu$
\begin{align}
\Re\Sigma(\omega) & \doteq  \frac {U^{2}n_{f}(1 - n_{f}) - 1}{\omega + \mu - (1 - n_{f})U } \  .
\end{align} 
From analyticity of the self-energy the pole in its real part induces a $\delta$-function singularity in its imaginary part at the critical point.

\section{Singularity in a two-particle irreducible vertex}

We investigate what impact  the singularity in the one-particle self-energy has on the behavior of two-particle functions. Since we stay within the mean-field approximation where all the generic (irreducible) functions are local we analyze the behavior of the local two-particle irreducible vertex.

\subsection{Two-particle approach}

We can derive an explicit representation of the general two-particle irreducible vertex via a generalized Ward identity.\cite{Janis01b} It reads
\begin{multline}
\label{eq:lambda-def}
\lambda(z_1,z_2) = \left. \frac{\delta\Sigma_{J}(z_1,z_2)}{\delta
    G_{J}(z_1,z_2)}\right|_{J=0} \\= \frac{1-a^{-1}(z_1,z_2)}{G(z_1)G(z_2)}
\end{multline}
where $\Sigma_{J}(z_1,z_2)$ and $G_{J}(z_1,z_2)$ are  the self-energy and the local Green function drifted  out of equilibrium by a weak external time-dependent source $J(t)$. Since the conduction electrons are subjected only to elastic, energy conserving scatterings, the two-particle vertex in equilibrium is described only by two independent frequencies. For the Falicov-Kimball model there are, however, two different irreducible vertices. That is, there are two ways how to assign two energy indices to four legs of the two-particle vertex.\cite{Janis10} One way corresponds to the fluctuations due to the Coulomb interaction, thermal averaging, and the other to the disorder-induced fluctuations, configurational averaging. The two different two-particle vertices do not intermingle.     We choose the latter two-particle vertex, since we do not need fluctuations in the density of the local $f$-electrons. We obtain for this disorder-induced vertex   
\begin{multline}\label{eq:a-def}
a(z_1,z_2) \\ =  \left\langle \prod_{\alpha=1}^{2}\frac{1}
{\left[1+G(z_\alpha)\left(\Sigma(z_\alpha)-V\right)\right]}\right\rangle_{V,T}\ .
\end{multline}

A pole or a singularity in irreducible vertex  $\lambda(z_1,z_2)$ can exist only at the real axis when the two frequencies coincide. An equation determining the pole is 
\begin{equation}\label{eq:lambdapole-full}
a(\omega \pm i0^{+}, \omega \pm i0^{+}) = \left\langle \frac{1}{\left[(\overline{\omega} - V)X \pm i s Y \right]^{2}}\right\rangle_{V,T} = 0 \ .
\end{equation}
When split into real and imaginary components we obtain two conditions in the notation of the preceding section
\begin{subequations}\label{eq:lambdapole}
\begin{multline}\label{eq:lambdapole-Re}
\hspace*{-10pt} \left\langle \frac{\overline{\omega} - V}{\left[(\overline{\omega} - V)^{2}X^{2} - s^{2} Y^{2} \right]^{2} + 4 s^{2}(\overline{\omega} - V)^{2}X^{2}Y^{2}}\right\rangle_{V,T} \\
 =0
\end{multline}
and
\begin{multline}\label{eq:lambdapole-Im}
\hspace*{-10pt} \left\langle \frac{(\overline{\omega} - V)^{2}X^{2}}{\left[(\overline{\omega} - V)^{2}X^{2} - s^{2} Y^{2} \right]^{2} + 4 s^{2}(\overline{\omega} - V)^{2}X^{2}Y^{2}}\right\rangle_{V,T} \\ = \left\langle \frac{s^{2} Y^{2}}{\left[(\overline{\omega} - V)^{2}X^{2} -  \right]^{2} + 4 s^{2}(\overline{\omega} - V)^{2}X^{2}Y^{2}}\right\rangle_{V,T}\ .
\end{multline}\end{subequations}
As in the case of the metal-insulator transition, equation~\eqref{eq:lambdapole-Re} determines energy $\overline{\omega}_{\lambda}$ at which the pole in the irreducible vertex occurs and Eq.~\eqref{eq:lambdapole-Im} determines the critical strength $V_{\lambda}$ of the random potential. 

To demonstrate the existence of the pole in the irreducible vertex we present a solution to Eqs.~\eqref{eq:lambdapole} in a homogeneous system without a static disorder. Equation~\eqref{eq:lambdapole-full} in this case reads
\begin{multline}\label{eq:a-F}
a_{\lambda} = \frac{1-n_f}{(1+G_\lambda\Sigma_\lambda)^2}\\ + \frac{n_f}{\left[1+G_\lambda(\Sigma_\lambda - U)\right]^2} = 0 \ .
\end{multline}
We introduced an abbreviation $F_\lambda=F(\omega_{\lambda})$ where $\omega_{\lambda}$ is the frequency at which irreducible vertex $\lambda(\omega_{\lambda} \pm i0^{+},\omega_{\lambda} \pm i0^{+})$ has a pole. Equation~\eqref{eq:a-F} can be rewritten to  
%
\begin{multline}
G_\lambda\left[U^2(n_f-1)G_\lambda-2U(n_f-1)(G_\lambda\Sigma_\lambda+1) \right. \\ \left. - \Sigma_\lambda(G_\lambda\Sigma_\lambda + 2)\right]=1 \ .
\end{multline}
This quadratic equation has two complex solutions for $G_{\lambda}$ 
\begin{equation}\label{eq:GF-critical}
G_\lambda=\frac{U(1 - n_{f}) - \Sigma_\lambda \pm i U \sqrt{n_f(1 - n_f)}}{\left[\Sigma_\lambda - U(1 - n_{f})\right]^{2} + U^{2}n_{f}(1 - n_{f})}\ .
\end{equation}
The Soven equation for the homogeneous Falicov-Kimball model, Eq.~\eqref{eq:SE}, can be rewritten to
\begin{align}\label{eq:Soven-homo}
G_{\lambda} &= \frac{n_{f} U - \Sigma_{\lambda}}{\Sigma_{\lambda}\left( \Sigma_{\lambda} - U\right)} \ .
\end{align}
Equaling the right-hand sides of Eqs.~\eqref{eq:GF-critical} and~\eqref{eq:Soven-homo} we obtain an equation for the self-energy
\begin{align}
\Sigma_{\lambda} = U\left[ n_{f}  \pm i \sqrt{n_{f}(1 - n_{f})}\right]\ .
\end{align}
The two roots stand for the advanced (positive imaginary part) and the retarded (negative imaginary part) self-energy. Inserting this self-energy in Eq.~\eqref{eq:Soven-homo} and using the Dyson equation~\eqref{eq:1PGF} we end up with a relation between interaction $U_{\lambda}$  and frequency $\omega_{\lambda}$ at the divergence of the irreducible vertex $\lambda(\omega_{\lambda} \pm i0^{+}, \omega_{\lambda} \pm i0^{+})$. If we set $\omega_{\lambda} = i0^{+}$ we then obtain an equation for an interaction $U_{\lambda}$ for which the pole in the irreducible vertex $\lambda$ reaches the Fermi energy. We show in the following subsection that the pole in vertex $\lambda$ can occur only at the Fermi energy.

\subsection{One-particle approach}

The pole in the two-particle irreducible vertex has at first glance no direct connection to the metal-insulator transition and the divergence in the one-particle self-energy. Derivation of the irreducible vertex via the functional derivative in Eq.~\eqref{eq:lambda-def} does not offer a quantitative relation to the self-energy.  The generalized Ward identity can, however, be resolved in special cases when the two frequency variables equal in the weak sense. We obtain for real frequencies two explicit algebraic relations between the self-energy and the irreducible vertex for real frequencies approached either from the opposite or the same complex half-planes.\cite{Janis10} The two-particle irreducible vertices then in the retarded-advanced and retarded-retarded channels read   
\begin{align}\label{eq:Ward+-}
  \lambda^{RA}(\omega,\omega) &= \frac{\Im\Sigma^{R}(\omega)}{\Im G^{R}(\omega)}= \frac 1 {\langle |G^{R}(\omega)|^{2}\rangle}\ , \\
\label{eq:Ward++}
\lambda^{RR}(\omega,\omega) &=\frac{Z(\omega)}{\langle G^{R}(\omega)^{2}\rangle}\ ,
\end{align}
where we denoted $Z(\omega)=\partial_\omega\Sigma^{R}(\omega)/(\partial_\omega\Sigma^{R}(\omega) - 1)$ and abbreviated $\partial_{\omega}  = \partial /\partial \omega$. We further introduced a notation
\begin{subequations}
\begin{align}
\langle G^R(\omega)^{2}\rangle &=\int_{-\infty}^\infty
\frac{d\epsilon\rho_0(\epsilon)}{(\omega + \mu - \Sigma^{R}(\omega)-\epsilon)^2}\nonumber \\
& = - \int_{-\infty}^\infty
\frac{d\epsilon\rho_0^{\prime}(\epsilon)}{\omega + \mu - \Sigma^{R}(\omega) - \epsilon}\ , \\
\langle \left|G^R(\omega)\right|^{2}\rangle &=\int_{-\infty}^\infty
\frac{d\epsilon\rho_0(\epsilon)}{|\omega + \mu - \Sigma^{R}(\omega)-\epsilon|^2} \ ,
\end{align}
\end{subequations}
where $\rho_{0}^{\prime}(\epsilon) = d\rho_{0}(\epsilon)/d \epsilon$. At the Fermi energy, $\omega = 0$ and the symmetric density of states, $\rho_{0}(-\epsilon) = \rho_{0}(\epsilon)$, we have $\Re\Sigma^{R}(0) = \mu$ and if $\mathrm{sign}(\rho_{0}^{\prime}(\epsilon)) = - \mathrm{sign}(\epsilon)$ then  $\langle G^R(0)^{2}\rangle < 0$\ .

It is evident that a pole in vertex $\lambda^{RA}$ can emerge only if $\Im G^{R}(\omega) = 0$. This happens just at the metal-insulator transition. It is different for vertex $\lambda^{RR}$.  Since the denominator on the right-hand side of Eq.~\eqref{eq:Ward++} is nonzero, a divergence in the irreducible vertex can be caused by a divergence in the scaling  factor $Z(\omega)$. It diverges if $\partial_{\omega}\Sigma^{R}(\omega) = 1$. The derivative of the self-energy off the Fermi energy ($\omega \neq 0$) is complex and it can never equal unity. The only chance to generate a divergent factor $Z(\omega)$ is at the Fermi energy. This happens just at an interaction $U_{\lambda}$.  We know that for the Fermi gas $Z(0) = 0$ and at the metal-insulator transition $\partial_{\omega} \Sigma^{R}(\omega)|_{\omega = 0} = \infty$. It is then clear that before the metal-insulator transition is reached, the irreducible vertex must go through a pole. That is, $U_{\lambda}< U_{c}$.  

The renormalization parameter $Z(\omega)$ changes sign at the point of divergence of vertex $\lambda^{RR}$.  The same happens with the local Green function. 
Since $\lambda^{RR}_F=\partial_\omega\Sigma_F/\partial_\omega G_F$, using Eq.~\eqref{eq:Ward++} we obtain
$\partial_\omega G_F=[\partial_\omega\Sigma_F-1]\langle G_F^2\rangle$, where $X_{F}$ stands for the value of function $X(\omega)$ at the Fermi energy. Then 
\begin{equation}
\label{eq:GderivC}
\partial_\omega \Re G_F=
\begin{cases}
> 0 & \mathrm{if}\quad U<U_\lambda \ ,  \\
<  0 & \mathrm{if}\quad U_\lambda<U<U_c.
\end{cases}
\end{equation}

The singularity in the local irreducible vertex $\lambda^{RR}$ does not induce a non-analytic behavior in physical, measurable quantities and is not connected with any symmetry-breaking transition. The full two-particle vertex in the mean-field limit, treated here, is 
\begin{multline}
\Gamma^{RR}(\omega,\omega;\mathbf{q}) =\\ \frac{\partial_{\omega}\Sigma^{R}(\omega)}{\partial_{\omega}\Sigma^{R}(\omega)\left(\left\langle G^{R}(\omega)^{2}\right\rangle - \chi^{RR}(\omega,\omega;\mathbf{q})\right) - \left\langle G^{R}(\omega)^{2}\right\rangle }
\end{multline}
where we denoted a two-particle bubble
\begin{equation}\label{eq:chi-bar}
 \chi^{RR}(\omega,\omega;\mathbf{q}) = \frac 1N\sum_{\mathbf{k}}
  G^{R}(\mathbf{k},\omega) G^{R}(\mathbf{k} + \mathbf{q},\omega)\ .
\end{equation}
At the Fermi energy vertex $\Gamma^{RR}(0,0;\mathbf{q})$ has no singularity and the divergence of vertex $\lambda^{RR}$ does not cause any violation of analytic properties of the full two-particle vertex and hence of the response function. This conclusion is in particular important for the calculation of vertex corrections to the mean-field physical quantities.\cite{Janis10,Pokorny13}

\subsection{Various densities of states}

We demonstrated that before the system reaches a critical metal-insulator transition it must go through a singularity in the two-particle irreducible vertex $\lambda^{RR}$. But a singularity in vertex $\lambda^{RR}$ need not indicate that the system experiences a metal-insulator transition.  In the preceding sections we demonstrated the existence of both singularities for a binary alloy, which is equivalent to the homogeneous Falicov-Kimball model with fixed densities of the local $f$-electrons.  We now determine the two divergencies for different densities of states of the homogeneous Falicov-Kimball model. We use dimensionless energy variables, this time multiples of the hopping amplitude $t$ between the nearest neighbors for the densities of states with infinite tails  or multiples of  the half-bandwidth $w$, when finite.   

\begin{itemize}
\item Lorentzian DOS.
\begin{equation}
\label{A1:DoS_lorentz}
\rho_0(\varepsilon)=\frac{1}{\pi}\frac{1}{\varepsilon^2+1} \ ,
\end{equation}
\begin{equation}
\label{A1:G_lorentz}
G(z)=\frac{1}{z+i\sigma}\ ,
\end{equation}
where $\sigma=\sgn(\Imm z)$. For the Falicov-Kimball model at half-filling we obtain 
$G_F=-4/(U^2+4)$. Notice that for this density of states $\left\langle \epsilon^{2}\right\rangle = \infty$ and hence no divergence for the self-energy can emerge, although vertex $\lambda^{RR}$ diverges at $U_{\lambda}=2$. 

\item Semi-elliptic DOS (infinite-dimensional Bethe lattice).
\begin{equation}
\label{A1:DoS_semi}
\rho_0(\varepsilon)=\frac{2}{\pi}\sqrt{1-\varepsilon^2}\ ,
\end{equation}
\begin{equation}
\label{A1:Gzero_semi}
G(z)=2z\left[1-\sqrt{1-\frac{1}{z^2}}\right] \ .
\end{equation}
For the Falicov-Kimball model at half-filling we obtain $G_F=-2\sqrt{1-U^2}$. The energy moments of this density of states can be calculated analytically
\begin{equation}
\left\langle \epsilon^{2n}\right\rangle = \frac 2\pi \int_{-1}^{1} d\epsilon \ \epsilon^{2n}\sqrt{1 - \epsilon^{2}} = \frac{(2n)!}{4^{n}n! (n + 1)!} \ .
\end{equation}
The relevant moments explicitly are $\left\langle \epsilon^{2}\right\rangle = 1/4$ and $\left\langle \epsilon^{4}\right\rangle = 1/8$. 

\item Gaussian DOS (infinite-dimensional hyper-cubic lattice).
\begin{equation}
\label{A1:DoS_gauss}
\rho_0(\varepsilon)=\pi^{-1/2}e^{-\varepsilon^2}\ ,
\end{equation}
\begin{equation}
G(z)=-i\sqrt{\pi}\wofz(z)\ ,
\end{equation}
where
\begin{equation}
\label{wofz}
\wofz(z)=\frac{i}{\pi}\int_{-\infty}^\infty dt \frac{e^{-t^2}}{z-t}
\end{equation}
is the Fadeeva error function. The energy moments of the Gaussian density of states are
\begin{equation}
\left\langle \epsilon^{2n}\right\rangle = \frac {1}{\sqrt{\pi}} \int_{-\infty}^{\infty} d\epsilon \ \epsilon^{2n}\exp\{- \epsilon^{2} \}  = \frac{(2n)!}{4^{n}n!}  \ .
\end{equation}
The two relevant moments explicitly are $\left\langle \epsilon^{2}\right\rangle = 1/4$ and $\left\langle \epsilon^{4}\right\rangle = 3/4$.

There is no  analytic expression for $G_F$, but we obtain an explicit equation  for $U_\lambda$ 
\begin{equation}
2\sqrt{\pi}x\exp(x^2)\erfc(x)=1\ , 
\end{equation}
where $x=U_\lambda/2$ with a  solution $U_\lambda\approx 0.8655$.
\\
\item Simple cubic/hyper-cubic (finite-dimensional) DOS.
\\
We assume the following normalization of the dispersion relation on the $d$-dimensional hyper-cubic lattice so that to allow for a meaningful limit to infinite lattice dimensions
$$
\epsilon(\mathbf{k}) = \frac t{\sqrt{d}}\sum_{i=1}^{d} \cos(k_{i}) \ 
$$
with the energy band on interval $(-\sqrt{d},\sqrt{d})$. 
The one-electron propagator is
\begin{equation}
\label{A1:3dSClGF}
G(z)=-i \sqrt{d} \int_0^\infty \!\! ds\  e^{i \sqrt{d} z s}J_0^d(s),
\end{equation}
where $J_0(t)$ is the Bessel function of first kind.
For $d=1,2$ and $3$ it can be reduced to a more useful form using complete elliptic integrals
of first kind $K(k)$.

The lowest energy moments for $d=2$ are $\left\langle \epsilon^{2}\right\rangle = 1/2$ and $\left\langle \epsilon^{4}\right\rangle = 9/16$. For $d=3$ they are $\left\langle \epsilon^{2}\right\rangle = 1/2$ and $\left\langle \epsilon^{4}\right\rangle = 5/8$.
\end{itemize}

For each of these densities of states we calculated interactions $U_{\lambda}$ and $U_{c}$ of the homogeneous Falicov-Kimball model at half filling with $n_{f} =1/2$. The results are summarized in Table~\ref{tab:DOS}. Notice that the critical interaction $U_{c}$ does not depend on the lattice dimension $d\ge 2$ and that the distance between $U_{\lambda}$ and $U_{c}$ systematically slightly increases with the chosen scaling of the hopping amplitude on the hyper-cubic lattices.  

\begin{table}
\begin{center}
\begin{tabular}{|c|c|c|c|}
\hline
DOS	&	
$w/t$	& 	
$U_\lambda/t$	& 	
$U_c/t$	\\
\hline
Lorentz			&	$\infty$		&	$2$					&	$\infty$		\\
semi-elliptic		&	$1$			&	$1/\sqrt{2}\approx 0.7071$	&	$1$			\\
square			&	$\sqrt{2}$	&	$0.8838$				&	$\sqrt{2}$		\\
simple cubic		&	$\sqrt{3}$	&	$0.9170$				&	$\sqrt{2}$		\\
$4d$ hypercubic	&	$2$			&	$0.8937$				&	$\sqrt{2}$		\\
$5d$ hypercubic	&	$\sqrt{5}$	&	$0.8908$				&	$\sqrt{2}$		\\
Gauss			&	$\infty$		&	$0.8655$				&	$\sqrt{2}$		\\
\hline	
\end{tabular}
\end{center}\caption{\label{tab:DOS} Different densities of states (DOS) with half-bandwidth $w$ and interactions $U_{\lambda}$ and $U_{c}$ for the half-filled homogeneous Falicov-Kimball model with the hopping amplitude between the nearest neighbors $t$. The value for $U_{\lambda}$ for the semi-elliptic DOS coincides with that obtained in Refs.~\onlinecite{Janis01a,Janis01b,Schafer13}.}
\end{table}

\section{Numerical results}

\begin{figure}[ht]
\begin{center}
\includegraphics[width=8cm]{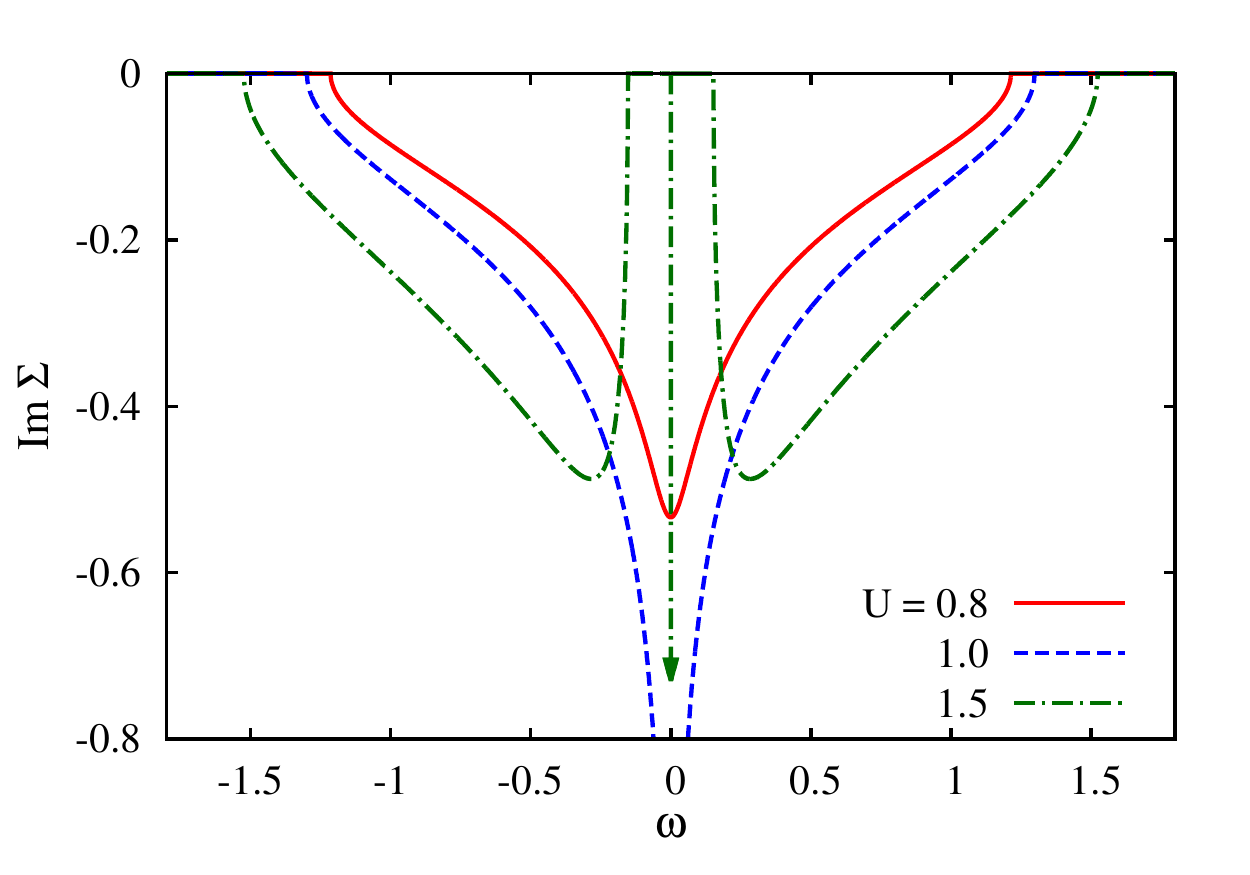}
\caption{(Color online) Imaginary part of the self-energy for FKM at half filling and interactions below ($U = 0.8$), at ($U=1$) and above ($U=1.5$)  the metal-insulator transition. Development of a divergence at the Fermi energy is apparent. \label{fig:ImSE-sym}}
\end{center}
\end{figure}

\begin{figure}[ht]
\begin{center}
\includegraphics[width=8cm]{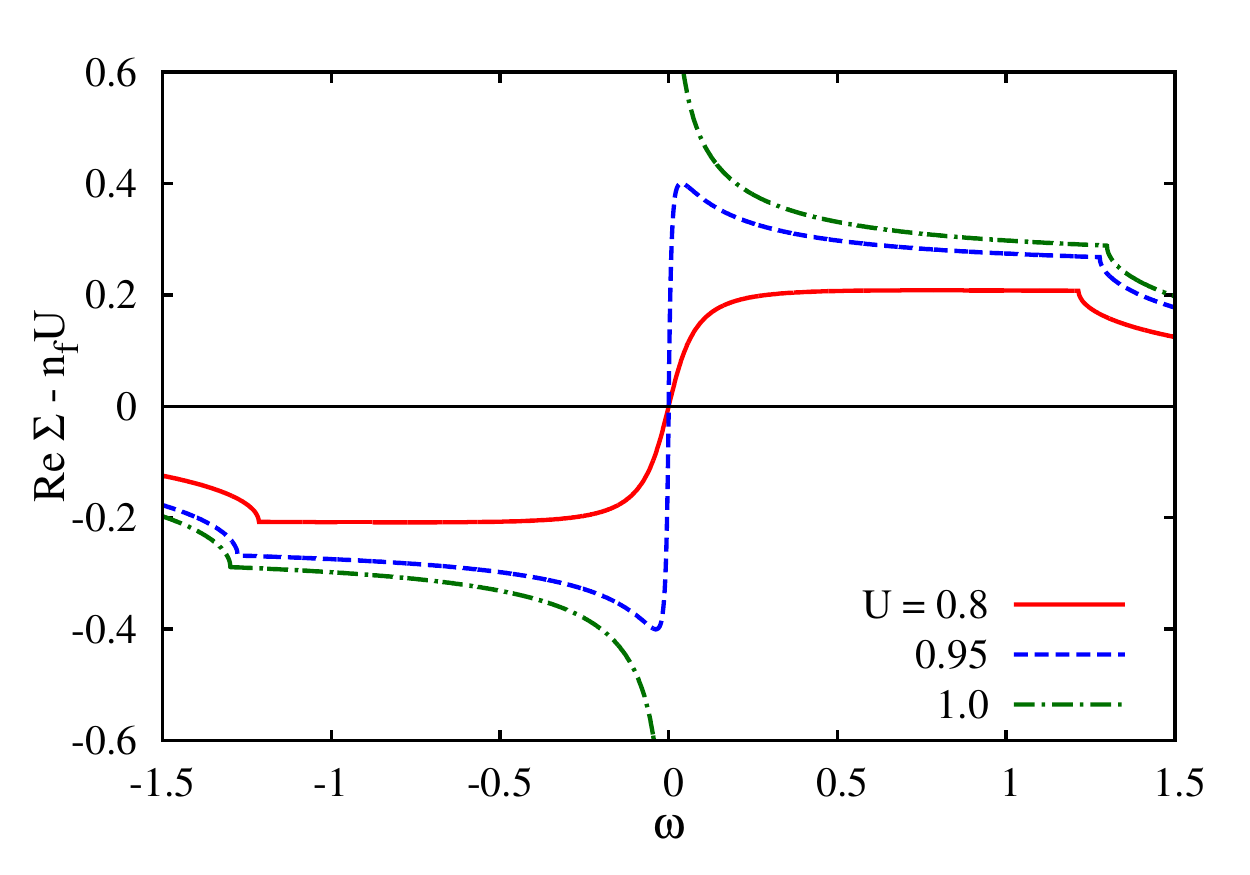}
\caption{(Color online) Real part of the self-energy for FKM at half filling for the same setting as in Fig.~\ref{fig:ImSE-sym}. \label{fig:ReSE-sym}}
\end{center}
\end{figure}

The exact solution of the Falicov-Kimball model in infinite dimensions allows us to compare the exact results with the predictions of the critical asymptotic expansion. Figs.~\ref{fig:ImSE-sym} and~\ref{fig:ReSE-sym} show the exact solution on an infinite-dimensional Bethe lattice (semi-elliptic DOS) for the imaginary and real parts of the self-energy near the metal-insulator transition at half filling. We can see development of a divergence in both parts of the self-energy. The real part of the self-energy starts to develop local extrema when approaching the critical point that reach the Fermi energy at the critical point.  The singularity at the Fermi energy survives also in the insulating state. 

\begin{figure}[ht]
\begin{center}
\includegraphics[width=8cm]{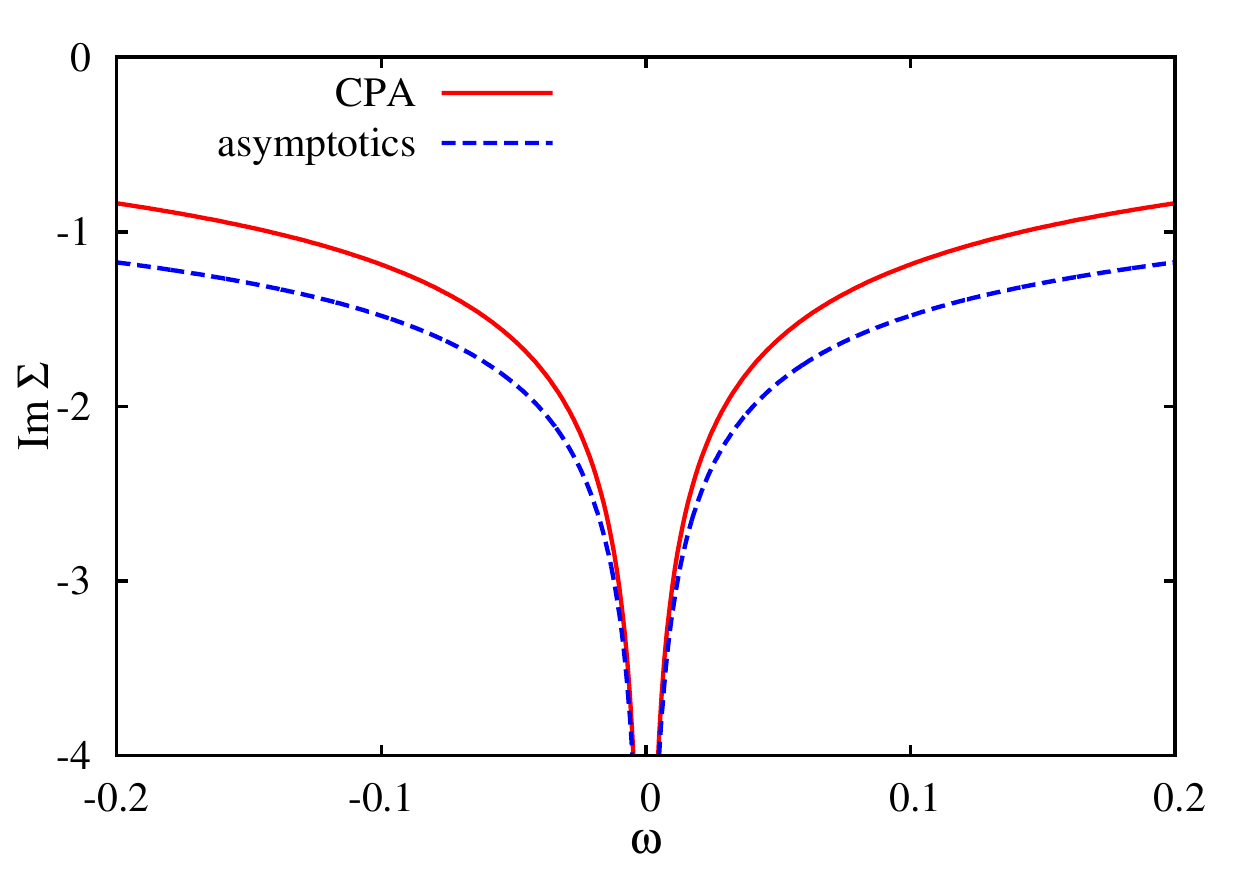}
\caption{(Color online) Comparison of the asymptotic and the exact (CPA)  imaginary part of the self-energy solutions near the critical point of FKM with a gaussian density of states. \label{fig:ImSE-asymp}}
\end{center}
\end{figure}

Our analysis of the critical behavior with a  diverging self-energy is universal and depends on the symmetric underlying density of states only via the second and fourth moments. It means that with the increasing strength of the scattering potential the self-energy develops a divergence for the symmetric distributions of the atomic potential independently of the underlying density of states. We demonstrated it explicitly in Fig.~\ref{fig:ImSE-asymp}. We used the gaussian DOS corresponding to the infinite-dimensional hyper-cubic lattice. We can see that the asymptotic solution at the critical point $U_{c}=\sqrt{2}$, Eq.~\eqref{eq:ImS-sym}, is very accurate.    

\begin{figure}[ht]
\begin{center}
\includegraphics[width=8cm]{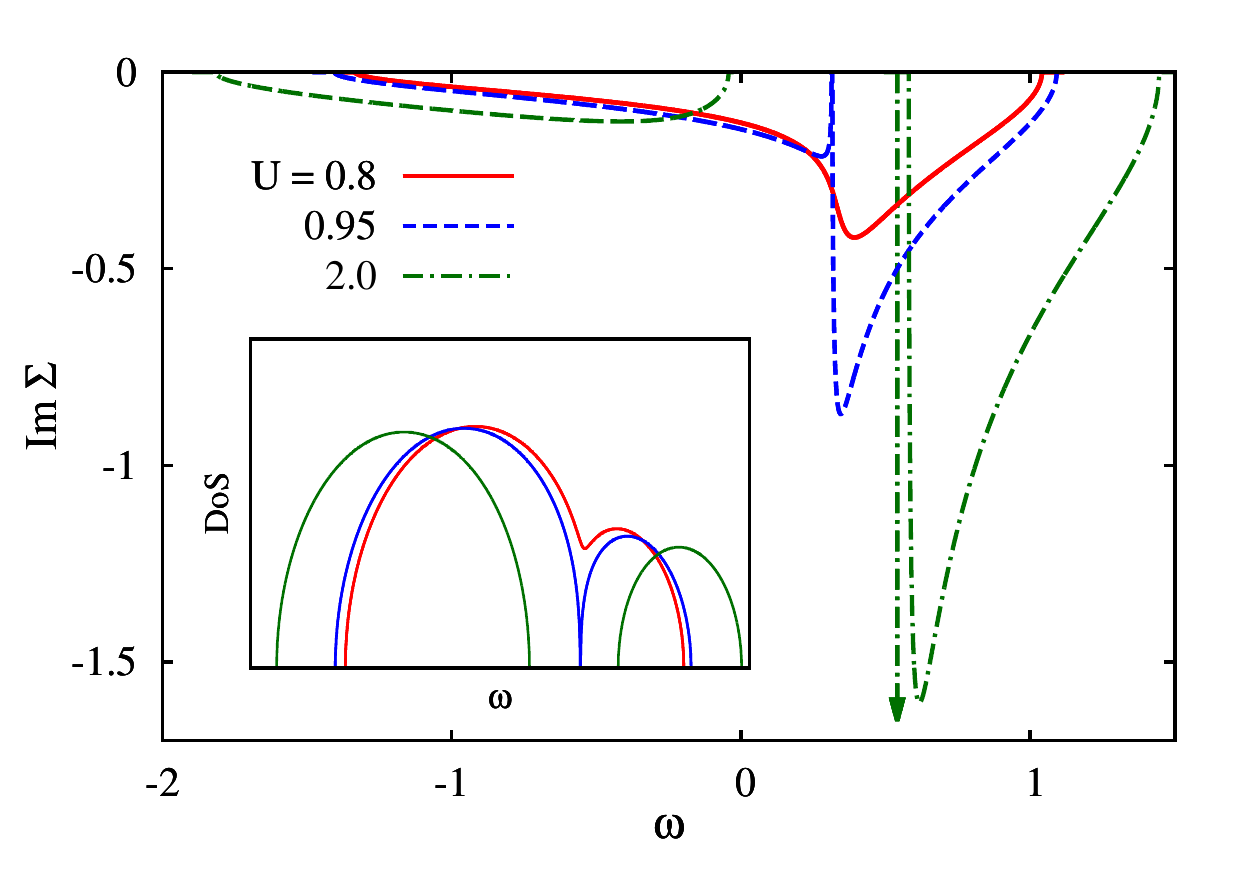}
\caption{(Color online) Imaginary part of the self-energy for an asymmetric distribution of the scattering potential of FKM with a fixed $n_{f} = 0.3$. The inset shows the corresponding spectral function. \label{fig:ImSE-asym}}
\end{center}
\end{figure}

\begin{figure}[ht]
\begin{center}
\includegraphics[width=8cm]{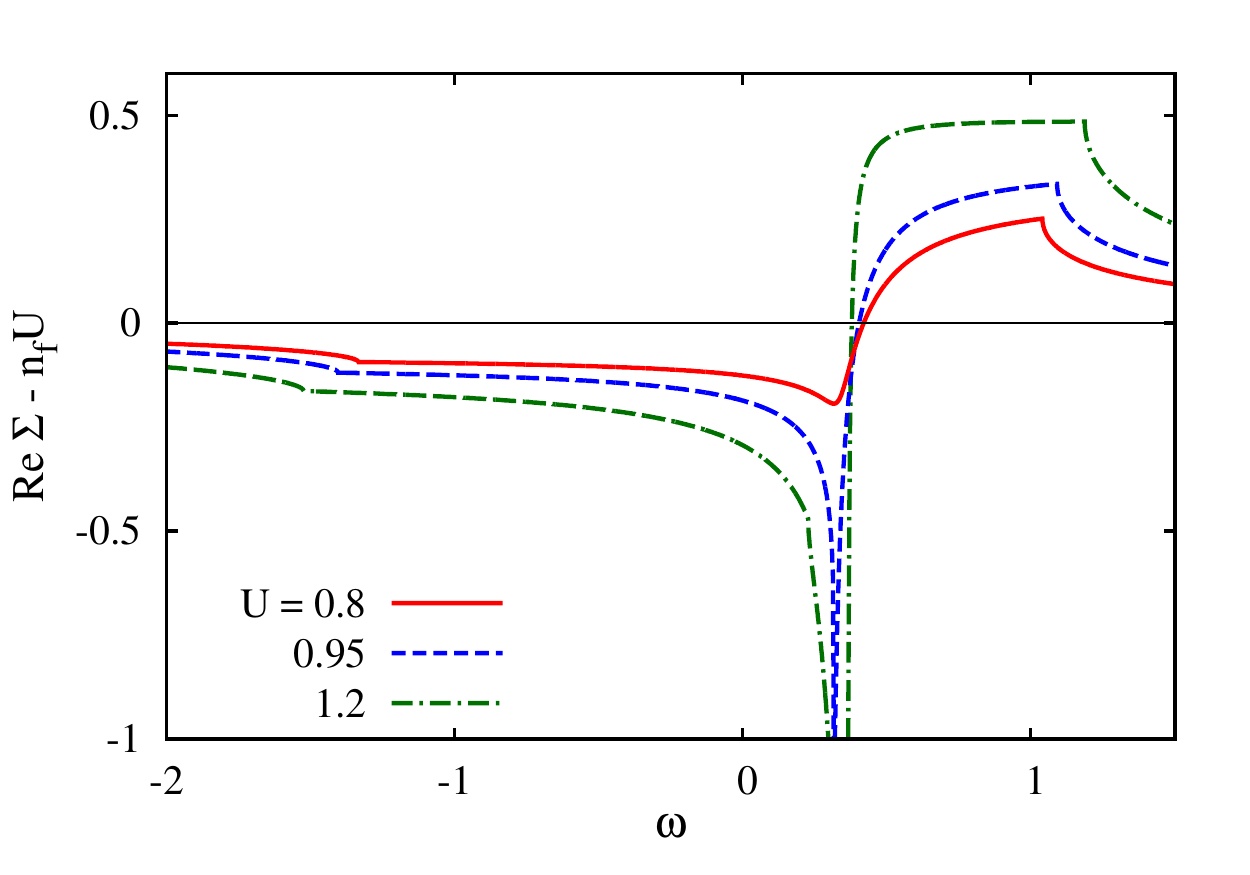}
\caption{(Color online) Real part of the self-energy for an asymmetric distribution of the scattering potential of FKM with a fixed $n_{f} = 0.3$.  \label{fig:ReSE-asym}}
\end{center}
\end{figure}

The split-band transition off the symmetric case (away from half filling of FKM) is generally non-critical. It means that also the metal-insulator transition is non-critical, except for a combination of the concentration of the scatterers and the strength of the scattering potential satisfying Eqs.~\eqref{eq:FKM-MIT1} and~\eqref{eq:FKM-MIT2} where the metal-insulator transition is critical with a diverging self-energy. We plotted the critical behavior of the imaginary part of the self-energy in  Fig.~\ref{fig:ImSE-asym} and of its real part in Fig.~\ref{fig:ReSE-asym} for $n_{f}=0.3$. We can see asymmetry in the behavior of the self-energy and that the critical transition is only at the internal edge of  the satellite band. It can also be seen how a pole in the real part of the self-energy peels off from the band edge in the insulating phase. The frequency asymptotics of this singularity is determined by Eq.~\eqref{eq:FKM-MIT1}. 

\begin{figure}[ht]
\begin{center}
\includegraphics[width=8cm]{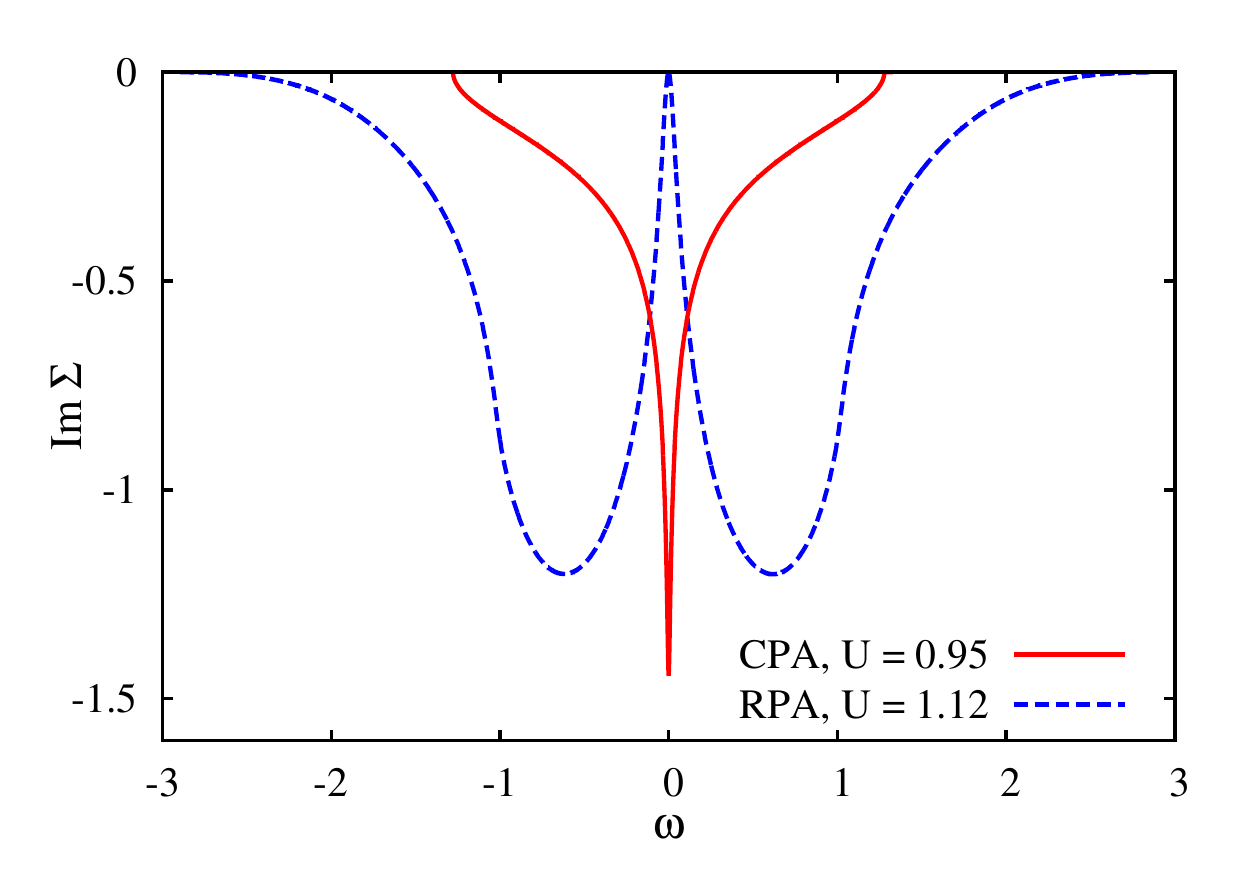}
\caption{(Color online) Imaginary part of the self-energy for FKM and the Hubbard model in a mean-field solution (RPA) at $U= 0.95 U_{c}$. Two maxima in $\Im\Sigma$ should reach the Fermi energy and close the rift at the metal-insulator transition. \label{fig:ImSE-comp} }
\end{center}
\end{figure}

\begin{figure}[ht]
\begin{center}
\includegraphics[width=8cm]{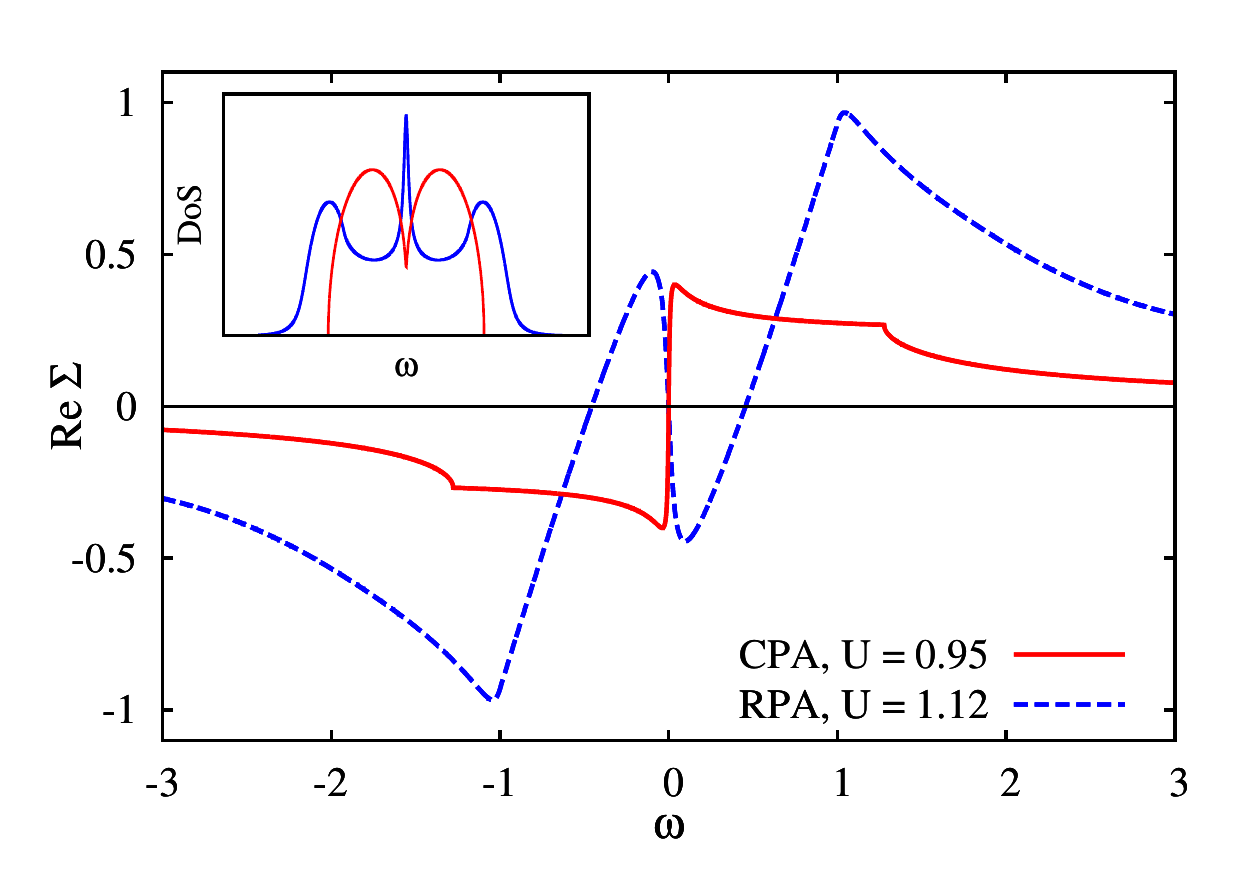}
\caption{(Color online) Real part of the self-energy near the metal-insulator transition of FKM compared with a mean-field solution of the Hubbard model (RPA) where the derivative at the Fermi energy have opposite sign. The inset shows the corresponding spectral function. \label{fig:ReSE-comp}}
\end{center}
\end{figure}

We analyzed the critical metal-insulator transition in the disordered Falicov-Kimball model. It is not a Fermi liquid and the imaginary part of the self-energy is non-zero at the Fermi energy. A physically more interesting case is the local Fermi liquid represented by the dynamical mean-field solution of the Hubbard model.  The self-energy vanishes at the Fermi energy for the Fermi liquid. Nevertheless, the imaginary part of the self-energy develops two peaks near the Fermi energy on the Kondo scale that should merge at the metal-insulator transition, destroying thereby the Fermi liquid and turning the metal a Mott insulator. A comparison of the imaginary part of the self-energy of the two models is plotted in Fig.~\ref{fig:ImSE-comp}. The solution of the Hubbard model used here, a non-self-consistent random-phase approximation (RPA) near the Hartree metal-insulator transition, does not contain the Kondo asymptotics and leads to a (critical) metal-insulator transition with a diverging self-energy. This transition is, however, unphysical. But more sophisticated numerical solutions such as the numerical renormalization group or the quantum Monte Carlo simulations support the existence of a metal insulator transition in the Hubbard model.\cite{Zhang93,Bulla99} It is, however, important, that  the behavior of the self-energy of the Hubbard model beyond the peaks in the imaginary part of the self-energy resembles that found in this paper near the Mott transition. The difference in the low-frequency behavior of the self-energy for the Fermi and non-Fermi liquid can be better seen in Fig.~\ref{fig:ReSE-comp} where the real part of the self-energy is plotted. The most important difference is the inverse gradient at the Fermi energy. The two extrema farther from the Fermi energy present the limits for the Fermi liquid properties and should diverge and reach the Fermi energy at the critical point. Unfortunately there is no analytic or semi-analytic solution that would show how the Fremi-liquid solution approaches the Mott transition.  

\begin{figure}[ht]
\begin{center}
\includegraphics[width=8cm]{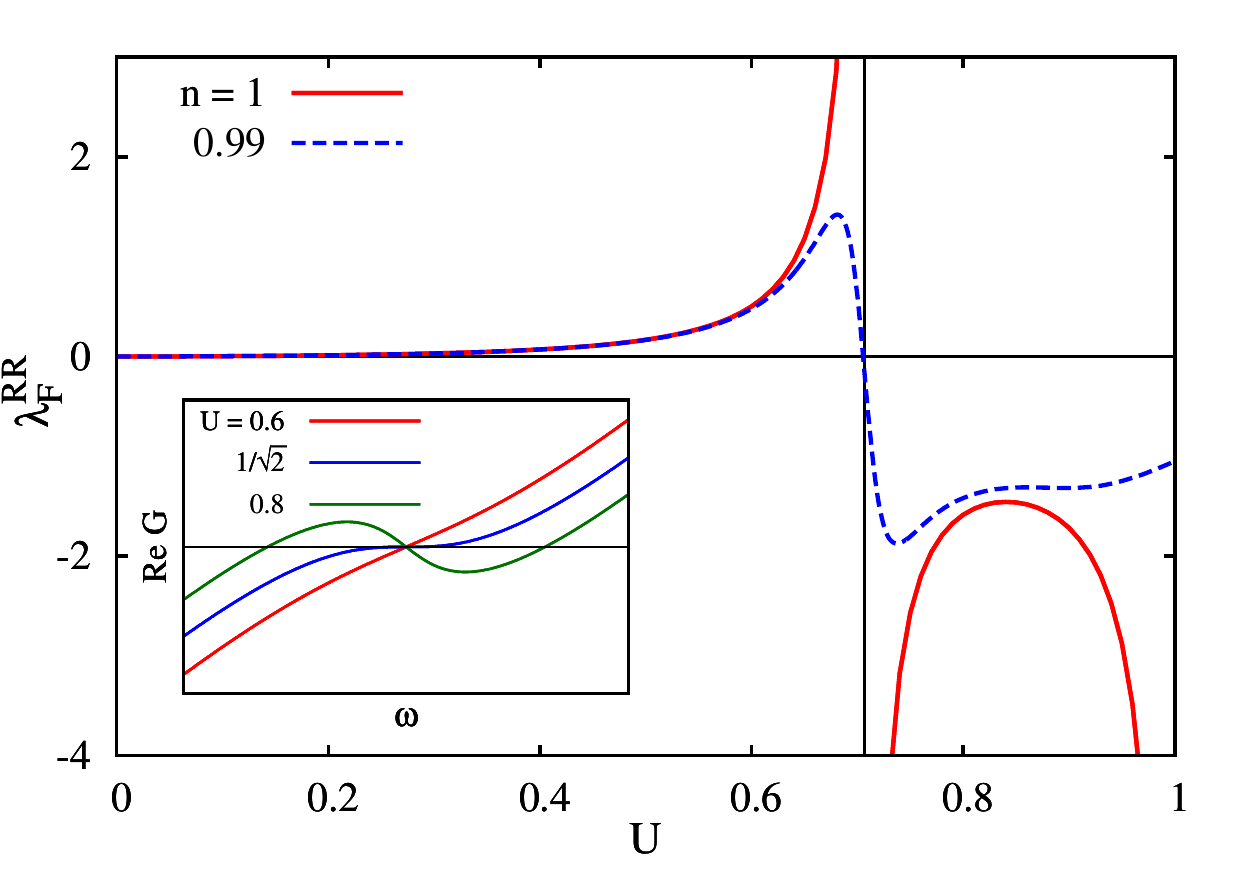} 
\caption{(Color online) Local irreducible vertex $\lambda^{RR}$
for the Falicov-Kimball model with the semi-elliptic DOS at half-filling
(solid red line) and slightly below half-filling (dashed blue line). The inset shows the corresponding real part of the one-electron propagator at half filling. \label{fig:lambda-comp}}
\end{center}
\end{figure}

We plotted the two-particle irreducible vertex $\lambda^{RR}$  in Fig.~\ref{fig:lambda-comp}. We can see how its derivative increases with approaching interaction $U_{\lambda}$ at which it  diverges. As discussed, this divergence is not connected with any symmetry breaking or transition to a new phase. It only leads to the change of the sign of the real part of the one-electron local propagator, cf. the inset of Fig.~\ref{fig:lambda-comp}.     

\begin{figure}[ht]
\begin{center}
\includegraphics[width=8cm]{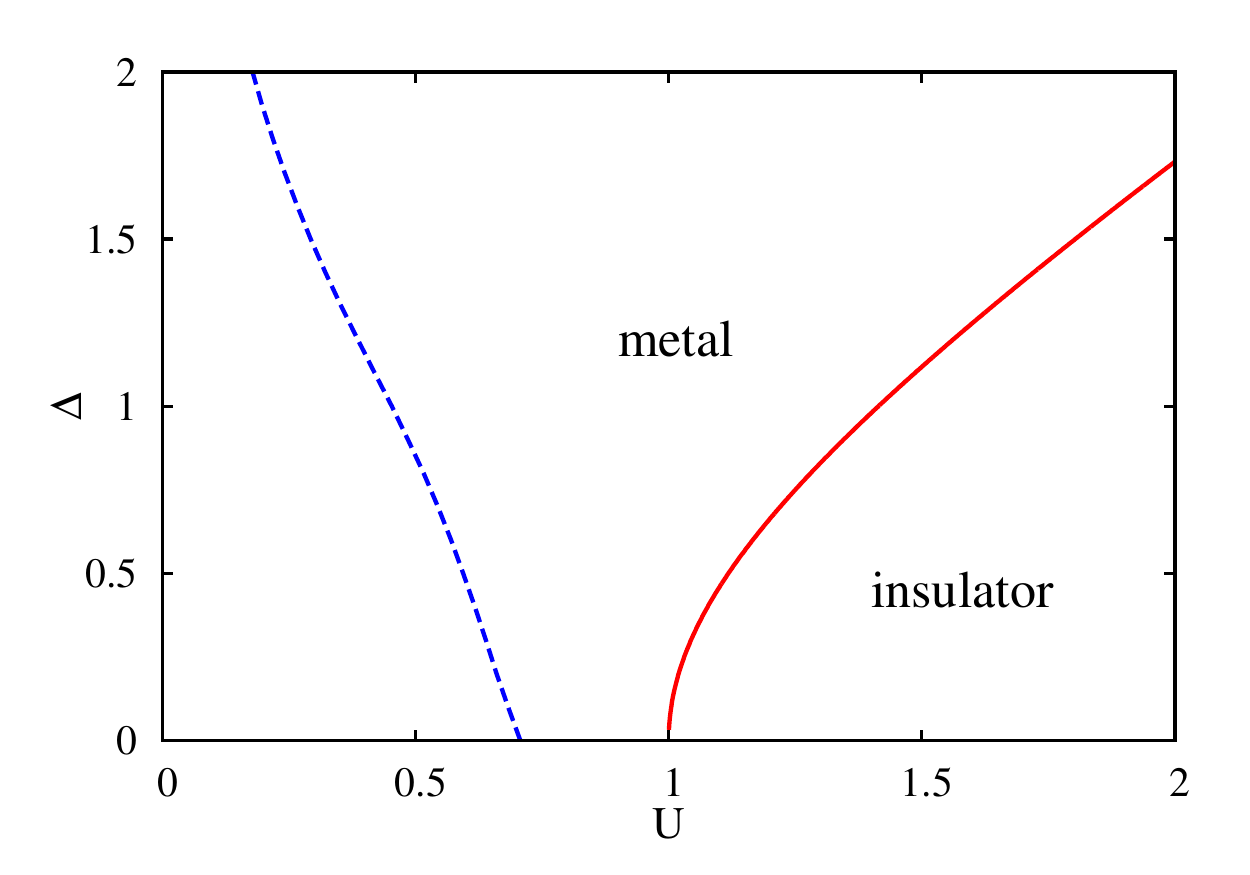}
\caption{(Color online) Phase diagram in the $U-\Delta$ plane for the disordered
Falicov-Kimball model with the box disorder. Solid red line: split-band (Mott) 
transition $\Delta_c=\sqrt{U^2-1}$. Dashed blue line: position of the pole in $\lambda^{RR}$ vertex. \label{fig:FKM-PT}}
\end{center}
\end{figure}

We argued that the critical metal-insulator transition is preceded by the divergence in vertex $\lambda^{RR}$. We plotted in Fig.~\ref{fig:FKM-PT} a phase diagram of a half-filled Falicov-Kimball model  with a box distribution of the random atomic potential indicating the lines of the two disorder strengths $\Delta_{\lambda}$ and $\Delta_{c}$ at which vertex $\lambda^{RR}$ and self-energy $\Sigma^{R}$ diverge, respectively. We used the following normalization of the box distribution \begin{equation}
\label{I:pdf_box}
\mathcal{P}(V)=\frac{1}{\Delta}\Theta\left[\left(\frac{\Delta}{2}\right)^2-V^2\right]=
\begin{cases}
 \frac{1}{\Delta}	& \text{for } |V| \leq \frac{\Delta}{2} \\
 0      	 		& \text{elsewhere}
\end{cases},
\end{equation}
where $\Theta$ is the Heaviside step function and $\Delta$ is the width of the distribution,
representing the disorder strength. It can be considered as the distribution of atomic energies of an infinite-component alloy.

\section{Discussion and conclusions}

We investigated the metal-insulator transition in disordered and interacting systems. We resorted to the interacting systems with only elastic scatterings of conduction electrons so that one can reach an explicit mean-field solution, the limit to infinite spatial dimensions. The metal-insulator transition can be driven by either electron correlations or by a configurational disorder or by their combined action. In the former case it is the Mott-Hubbard transition and in the latter one it is the split-band transition. The two transitions share a number of common features. The former transition is marked by the diverging self-energy at the Fermi energy. The latter one can either be non-critical or critical. The self-energy at the non-critical transition is continuous at the transition point. Such a transition occurs when the Fermi energy reaches an edge of the renormalized energy band. This is a transition from a metal to a band insulator. The metal-insulator transition in disordered systems may also be induced by a divergence in the self-energy. We manifested in this paper that the critical disorder-driven metal-insulator transition shares its universal critical behavior with the interaction-driven Mott transition.  

We assumed the metallic phase and a divergence of the imaginary part of the self-energy and introduced an asymptotic  perturbation expansion in the critical region of the metal-insulator transition for the disordered Falicov-Kimball model within the dynamical mean-field approximation. We derived the leading low-energy asymptotics for the real and imaginary parts of the self-energy for arbitrary densities of states and arbitrary distributions of the random potential. We showed that for the symmetric distributions of randomness, $P(V) = P(-V)$, both imaginary and real parts of the self-energy display the same singularity of order $\Delta\omega^{-1/3}$ at the critical point $\omega_{0}$ with $\Delta\omega = \omega - \omega_{0}$. For the asymmetric distributions the imaginary part of the self-energy  diverges at the critical point as $\Delta\omega^{-1/4}$  while the real part of the self-energy remains bounded from the metallic side. It has  a square-root singularity from the insulating side of the transition where the imaginary part of the self-energy is zero. The diverging self-energy for the asymmetric distributions of the scattering potential does not occur at the split-band limit but later at the internal band edge of the satellite band separated from the central one by an energy gap. 

The expansion used to derive the critical behavior at the metal-insulator transition is justified only for bounded energy bands of the conduction electrons. For energy bands with infinite tails the critical point with  the diverging self-energy, vanishing of the spectral function, exists but there is no insulating phase. The gap states are filled by the electrons from the long tails of the unperturbed density of states. The strong-scattering limit then corresponds to a semi-metal. To describe properly the strong-scattering regime in systems with infinite band tails one needs to supplement the expansion in small parameters by a contribution from the pole of the one-electron Green function, Eq.~\eqref{eq:exp-tails}. The addition of the contributions from the long tails of the bare density of states does not, however, affect the critical behavior derived here. 

The presented analysis of the metal-insulator transition directly applies only to non-Fermi-liquid systems. In the homogeneous Hubbard model treated within DMFT the imaginary part of the self-energy vanishes at the Fermi energy and the derivative of its real part has negative sign at the Fermi energy, that is $\partial_{\omega} \Sigma_{F} < 0$. Thus, it  is more complicated to prove the existence of the Mott transition in the Hubbard model.  A reliable analytic solution displaying the metal-insulator transition in the Hubbard model  does not exist, but numerical studies indicate that the imaginary part of the self-energy reaches high maxima at the Kondo scale around the Fermi energy beyond which the behavior of the self-energy is very much analogous to that studied here. The metal-insulator transition cannot be reached if the Fermi-liquid low-energy asymptotics dominates such as is the case of the single-impurity Anderson model. If the impact of the low-frequency Fermi-liquid asymptotics can be suppressed by the additional self-consistency of the dynamical mean-field theory, the critical behavior of the self-energy at the Mott-Hubbard transition will be the same as we derived here. That is, should the Kondo scale shrink to zero, the two peaks of the imaginary part of the self-energy would reach the Fermi energy and the self-energy would display the same singular behavior as we found in the Falicov-Kimball model. In this respect our findings about the critical behavior of the metal-insulator transition are universal. What still remains to prove is whether and how fast the Kondo scale may vanish at a finite interaction strength. 
         
Apart from the singularity connected with the metal-insulator transition we addressed the existence of a pole in a local two-particle irreducible vertex and its relation to the critical metal-insulator transition. We found that a pole in the local irreducible vertex $\lambda^{RR}$ exists  at the Fermi energy in all models showing the metal-insulator transition. Divergence in vertex  $\lambda^{RR}$ precedes the metal-insulator transition, but this singularity need not indicate the existence of the metal-insulator transition. A divergence in vertex $\lambda^{RR}$ occurs when $\partial_{\omega} \Sigma_{F} = 1$ while  $\partial_{\omega} \Sigma_{F} = \infty$ at the metal-insulator transition. We demonstrated that the singularity in  vertex $\lambda^{RR}$ is not connected with any symmetry breaking and does not cause non-analytic behavior in the measurable response functions. We concluded that to avoid unphysical singularities in the expansion beyond the local dynamical mean-field approximation, it is necessary to formulate the expansion in such a way that instead of the local irreducible vertex $\lambda$ one uses the fully  local vertex $\gamma(z_{1},z_{2})=\lambda(z_{1},z_{2})/(1 - \lambda(z_{1},z_{2}) G(z_{1}) G(z_{2}))$ of the dynamical mean-field theory where all the one-electron propagators are replaced by their local elements. Such an expansion is free of any unphysical singularity and no cancellation rules should be controlled.\cite{Pokorny13} This conclusion is important for the expansions beyond the dynamical mean-field theory that include non-local dynamical fluctuations via two-particle vertices such as the parquet equations,\cite{Janis01b} asymptotic solution in high spatial dimensions,\cite{Janis05a,Janis05b} the dual-fermion construction,\cite{Rubtsov07}, the dynamical-vertex approximation, \cite{Toschi07}   one-particle irreducible functional approach,\cite{Rohringer13} or multi-scale extensions of quantum clusters method.\cite{Slezak09}  

\section*{Acknowledgment}
 
 We thank J. Koloren\v c for fruitful and inspiring discussions.

\end{document}